# Combining Beamforming and Space-Time Coding Using Noisy Quantized Feedback

Siavash Ekbatani[(*)] IEEE Student Member  and  Hamid Jafarkhani IEEE Fellow

## Abstract

The goal of combining beamforming and space-time coding in this work is to obtain full-diversity order and to provide additional received power (array gain) compared to conventional space-time codes. In our system, we consider a quasi-static fading environment and we incorporate both high-rate and low-rate feedback channels with possible feedback errors. To utilize feedback information, a class of code constellations is proposed, inspired from orthogonal designs and precoded space-time block codes, which is called generalized partly orthogonal designs or generalized PODs. Furthermore, to model feedback errors, we assume that the feedback bits go through binary symmetric channels (BSCs). Two cases are studied: first, when the BSC bit error probability is known a priori to the transmission ends and second, when it is not known exactly. In the first case, we derive a minimum pairwise error probability (PEP) design criterion for generalized PODs. Then we design the quantizer for the erroneous feedback channel and the precoder codebook of PODs based on this criterion. The quantization scheme in our system is a channel optimized vector quantizer (COVQ). In the second case, the design of the quantizer and the precoder codebook is based on similar approaches, however with a worst-case design strategy. The attractive property of our combining scheme is that it converges to conventional space-time coding with low-rate and erroneous feedback and to directional beamforming with high-rate and error-free feedback. This scheme shows desirable robustness against feedback channel modeling mismatch.

This work was supported in part by an ARO Multi-University Research Initiative (MURI) grant number # W911NF-04-1-0224. The authors are with the center for pervasive communications and computing (CPCC) at the department of electrical engineering and computer science (EECS), University of California, Irvine (UCI). Email: `sekbatan@uci.edu` and `hamidj@uci.edu`. [(*)]Address: 3400 Calit2, # 19, Irvine, CA, 92697-2800. Phone: +1(949)824-3020.

# I. INTRODUCTION

We study point-to-point closed-loop communications over multiple-input single-output (MISO) quasi-static Rayleigh fading channels. This channel model characterizes the downlink of several wireless communication systems such as the new generation of fixed and mobile cellular systems in IEEE 802.16 [1]. We assume that the single-antenna receiver is able to estimate the channel coefficients perfectly. Acquiring perfect receiver channel state information (CSI) is relatively easy. Because by definition, in quasi-static channel environments, there is a long fading block length and accurate channel estimation through training is possible. Obtaining transmitter CSI (CSIT), however, requires the use of feedback in the system. When CSIT is available either fully or partially, directional beamforming can increase the average received signal-to-noise ratio (SNR). This benefit of beamforming is called array gain. It improves the capacity of wireless communication systems [2]–[4], reduces the outage probability [5] [6], and enhances the error performance. Partial CSIT is widely used in the literature for combining space-time coding and beamforming using precoded space-time codes, [1], [7]- [11], and for combining beamforming and power control [12] [13].

We design a fixed-rate practical coding and beamforming system with instantaneous power constraint at the transmitter. The performance measure is the pairwise error probability (PEP) of transmission over a finite length data frame. The diversity order in this context is defined as the decay rate of the pairwise error probability with SNR. In MISO systems, full-diversity order means an error probability decay rate equal to the number of transmit antennas. In quasi-static environments, channel fading coefficients remain constant during the transmission of each data frame, which occurs within the channel coherence time, and change independently afterwards. Therefore, feedback information must be updated in every frame. Usually, feedback channels are severely bandwidth and power limited. It is preferable to reduce the feedback rate, using resolution-constrained (quantized) CSI [14]- [19]. Most of the work in the literature assume error-free (noiseless) feedback. However, this assumption requires tremendous protection of feedback information in a practical system, which requires the dedication of excessive bandwidth and power to feedback channels. We study a practically motivated scenario, where not only is the feedback link resolution constrained, but also the existence of noise in the feedback link introduces errors to feedback information [20]. This situation can naturally arise in a variety of applications, where feedback information must be carried through actual fading channels [21]. If the system is designed robustly against feedback errors, a significant portion of the additional bandwidth and excessive power spent for protecting feedback information can be saved. This can increase the efficiency of the overall network.

For simplicity, we consider a simple discrete memoryless channel (DMC) model of the feedback link, using binary symmetric channel (BSC) models for feedback bits. We assume that the cross-over probability (bit error probability) of feedback bits is fixed over all channel realizations. This is a simple, yet practical version of the bit error model in [22], which introduces a finite-state fading channel model. A more elaborate model of feedback errors can be found in [23]. Other than the limited bandwidth and feedback errors, possible delay in feedback can also be problematic. With delay, feedback information might become outdated. Therefore, the system performance degrades due to the mismatch between the CSIT and the real channel realization [13] [24]. In this work, however, we only consider a simple erroneous feedback channel model to establish the design guidelines and leave the extensions to other feedback channel models for future work.

If feedback information is error-free, beamforming or combining space-time coding and beamforming can provide full-diversity order on top of array gain [25] [26]. However, with feedback errors, achieving full-diversity order is not straightforward [20] [21]. It was recently shown that with less restriction on the feedback rate, we can obtain full-diversity order by combining space-time coding and beamforming [25]- [32]. This is even possible with feedback errors in some scenarios. For example in [33] this issue is addressed by appropriate spatial power allocation using mean and covariance feedback in block-fading channels. In our combining scheme in this work, the goal is to address the low-rate feedback and feedback error problems at the same time for quasi-static channels. Moreover, we want our scheme to extend easily for any feedback rate and feedback error probability, even if there is a mismatch between the actual feedback channel parameters and the knowledge of the transmitter/receiver sides about these parameters.

The rest of the paper is organized as follows: In Section II, the system model is sketched and our combining scheme is introduced. This scheme is based on a code structure named generalized partly orthogonal designs (PODs). Then we derive the design criteria of our combining scheme, based on PEP analysis. In Section III, using channel optimized vector quantizers (COVQs) similar to [35]- [38], we propose a quantizer structure for the feedback channel and a precoder structure for generalized PODs. First we consider the case when feedback error probability is known. The case of feedback channel modeling mismatch, or not knowing feedback error probability a priori is also discussed. The latter discussion is inspired from the worst-case COVQ design approaches in [39]. Section IV provides the numerical results of the paper and finally Section V draws our major conclusions.

*Notations*: In the sequel, $\mathrm{p}(i)$, $\mathrm{p}(j|i)$, and $\mathrm{p}(j;i)$ denote the marginal probability of the event $i$, the conditional probability of the event $j$ given $i$, and the joint probability of events $j$ and $i$, respectively. The operators $\| \cdot \|_F$, $(\cdot)^T$, $(\cdot)^\dagger$, and $(\cdot)^*$ represent Frobenius norm, transpose, Hermitian, and conjugate, respectively. $\mathrm{f}(\mathrm{x})$ denotes the probability density function (pdf) of the random vector $\mathrm{x}$ and $\mathrm{f}(\mathrm{x}|i)$ denotes the pdf of $\mathrm{x}$ given event $i$. Finally, $\mathbb{E}_\mathrm{x}\{\Psi(\mathrm{x})\}$ shows the expectation of a function of $\mathrm{x}$ and $\mathbb{E}_{\mathrm{x}|i}\{\Psi(\mathrm{x})\}$ shows the same expectation given event $i$.

## II. SYSTEM COMPONENTS AND DESIGN CRITERIA

The block diagram of our system is shown in Fig. 1. In this system, the receiver is equipped with perfect channel estimation, ideal synchronization, and maximum-likelihood (ML) decoding.

### A. Forward and Feedback Channel Parameters

During the transmission of each data frame, we represent the MISO quasi-static fading forward channel coefficients by an $M$-dimensional complex Gaussian vector $\widehat{\mathbf{h}}$. The task of the quantizer is to quantize this vector at the receiver and generate a quantization index. The feedback channel conveys the quantization index back to the transmitter. Due to the limited feedback bandwidth, in some cases we quantize a portion of $\widehat{\mathbf{h}}$ by decomposing this vector into two parts as

$$\widehat{\mathbf{h}} = \left[ \widetilde{\mathbf{h}}^T_{(M-N)\times 1} \quad \overline{\mathbf{h}}^T_{N\times 1} \right]^T \tag{1}$$

where $\overline{\mathbf{h}}$ is the $N$-dimensional quantized part and $\widetilde{\mathbf{h}}$ is the $(M - N)$-dimensional part which remains unknown to the transmitter. The parameter $N$ in our system that is the dimension of the quantized vector depends on the number of feedback indices in the feedback codebook. Note that we deal with short-term (instantaneous) power constraint at the transmitter. Also our scheme only supports fixed-rate transmissions. Therefore, the amplitude of the quantized channel vector is irrelevant to the performance and the design of our system. Our quantization scheme works on the direction of the channel sub-vector $\overline{\mathbf{h}}$, i.e., $\overline{h} = \overline{\mathbf{h}}/\|\overline{\mathbf{h}}\|_F$. We decompose $\overline{\mathbf{h}}$ into its amplitude and its direction through $\overline{\mathbf{h}} = \sqrt{\gamma}\, \overline{h}$, where $\gamma$ is the amplitude square. The direction vector $\overline{h}$ is uniformly distributed over the unit amplitude complex sphere $\mathcal{C}^N$ and the amplitude square $\gamma$ is a Chi-square random variable with $2N$ degrees of freedom. The pdf of $\gamma$ is $\mathrm{f}(\gamma) = \gamma^{N-1}\exp{(-\gamma)}/(N-1)!$ .

In our feedback scheme, we partition the space of $\overline{h}$ into $K$ mutually exclusive feedback (Voronoi) regions $\mathcal{V} = \{\mathcal{V}_1, \cdots, \mathcal{V}_K\}$, whose union spans the whole unit-norm complex sphere $\mathcal{C}^N$. The objective of the quantizer at the receiver is to encode $\overline{h}$ by finding the index of the Voronoi region $\mathcal{V}_i$ that $\overline{h}$ belongs to. The objective at the transmitter is to choose a precoder matrix $\mathcal{P}_j$ based on the transmitter's feedback index $j$, which may be different from $i$ as a result of errors in the feedback channel. We first need to design a precoder matrix codebook $\mathcal{P} = \{\mathcal{P}_1, \cdots, \mathcal{P}_K\}$ and store it at both ends of the communication link. The precoder matrix $\mathcal{P}_j$ defines a matrix constellation set at the transmitter.

In our system, after finding a feedback index, the receiver converts it to bits and sends the bits to the transmitter. The typical value of the feedback bit error probability is approximately $0.04$ [20] [21]. However, for the sake of generality, we will consider a case where feedback bit error, $\rho_f$, can take any value between $0.00$ and $0.50$. Note that if $\rho_f > 0.50$, the transmitter can simply flip the feedback bits and achieve better performance results. In the sequel we assume that the transmitter/receiver sides of our system know $\rho_f$ a priori, unless stated otherwise. Suppose that the feedback index $i \in \{1, \cdots, K\}$ is chosen at the receiver. The probability of this event is denoted by $\mathrm{p}(i)$. Also, suppose that the mapping of feedback indices to feedback bits is an identity mapping. For any feedback index $i$, the feedback bits can be simply obtained as binary representation of $i - 1$. The encoder of the quantizer (and the mapper) are located at the receiver and the decoder of the quantizer (and the demapper) are implemented using a simple table look-up operation at the transmitter. In our system model, feedback indices go through a DMC from $\log_2 K$ uses of BSCs with cross-over error probabilities $\rho_f$ for each feedback bit. The feedback channel converts $i$ to $j \in \{1, \cdots, K\}$, with the conditional index inversion probability

$$\mathrm{p_f}(j|i) = (\rho_f)^{d(i,j)} \left(1 - \rho_f\right)^{\log_2 K - d(i,j)} \tag{2}$$

where $d(i, j)$ is the Hamming distance between the binary representations of $j$ and $i$.

### B. Modulation Scheme

The transmitter uses a class of coded modulation schemes called generalized partly orthogonal designs (PODs). PODs were firstly introduced in [40]. Using PODs, we can utilize feedback information from any number of channel coefficients $N$ for combining coding and beamforming across $M$ transmit antennas, where $M \geq N$. The precoder or beamformer structures incorporated in original PODs of [40] were designed based on maximizing the received SNR. The generalized PODs of this work are however designed based on minimizing the pairwise error probability and hence they have a different structure. Unlike the original PODs, the precoder structures used in generalized PODs allow spatial power allocation, depending on the probability of error in the feedback link. Like original PODs, generalized PODs also have two parts: i) the STBC inner code that uses an $M \times T$ orthogonal design from [41] [42] and ii) the precoding part, which uses an $N \times N$ ($N$-dimensional) precoder matrix, chosen based on feedback information. Each modulation matrix in the constellation set is limited to a short-term (instantaneous) power constraint $MT$.

For clarification, let us use the following example. Suppose that the MISO forward channel has 4 transmit antennas. Similar to [40], we construct the code structures on the columns of an orthogonal design matrix. With a slight abuse of the notation, in this example we use the following

$M \times T$ orthogonal design structure where $M = T = 4$

$$\mathcal{Z} = \begin{bmatrix} z_1 & -z_2 & -z_3 & -z_4 \\ z_2 & z_1 & z_4 & -z_3 \\ z_3 & -z_4 & z_1 & z_2 \\ z_4 & z_3 & -z_2 & z_1 \end{bmatrix}$$

Here, $z_\kappa : \kappa \in \{1, \cdots, T\}$ denote the data symbols chosen from a real symbol constellation. Generalized PODs allow using any precoder dimension $N$, where $N \leq M$. The dimension of the precoder depends on the dimension of the channel vector quantizer. For instance, using a 2-dimensional precoder, combining is performed through the following code structure:

$$\mathcal{Z}_j^2 = \begin{bmatrix} z_1 & -z_2 & -z_3 & -z_4 \\ z_2 & z_1 & z_4 & -z_3 \\ \mathcal{P}_j^{2\times2}\begin{pmatrix} z_3 \\ z_4 \end{pmatrix} & \mathcal{P}_j^{2\times2}\begin{pmatrix} -z_4 \\ z_3 \end{pmatrix} & \mathcal{P}_j^{2\times2}\begin{pmatrix} z_1 \\ -z_2 \end{pmatrix} & \mathcal{P}_j^{2\times2}\begin{pmatrix} z_2 \\ z_1 \end{pmatrix} \end{bmatrix} \tag{3}$$

Also, using a 4-dimensional precoder, we can use the following code:

$$\mathcal{Z}_j^4 = \begin{bmatrix} \mathcal{P}_j^{4\times4}\begin{pmatrix} z_1 \\ z_2 \\ z_3 \\ z_4 \end{pmatrix} & \mathcal{P}_j^{4\times4}\begin{pmatrix} -z_2 \\ z_1 \\ -z_4 \\ z_3 \end{pmatrix} & \mathcal{P}_j^{4\times4}\begin{pmatrix} -z_3 \\ z_4 \\ z_1 \\ -z_2 \end{pmatrix} & \mathcal{P}_j^{4\times4}\begin{pmatrix} -z_4 \\ -z_3 \\ z_2 \\ z_1 \end{pmatrix} \end{bmatrix} \tag{4}$$

In these codes, $\mathcal{P}_j^{N\times N}$ denotes an $N \times N$ precoder matrix with power $N$, i.e., $\text{Tr}\left( \mathcal{P}_j^{N\times N} \mathcal{P}_j^{N\times N\dagger} \right) = N$, where $\text{Tr}(\cdot)$ denotes the trace operation. Note that when $N = M$ generalized PODs look similar to a conventional PSTBC in [25]. Also when $M \neq N$, they look like a POD in [40]. However, they are different structures, as we will clarify further in the sequel. Using similar approaches, one can design generalized PODs from complex orthogonal designs for any dimensions [42]. Furthermore, similar ideas can be applied to design a generalized partly quasi-orthogonal design based on quasi-orthogonal inner STBCs from [43] [44]. The main challenge is how to design the precoder matrix $\mathcal{P}_j^{N\times N}$ to obtain minimum pairwise error probability with each code structure.

Suppose that the quantizer codebook has $K$ indices. We will show that in order to obtain full-diversity order using the above codes, we must choose precoder structures with dimension $N \leq K$. If $K \geq M$, we have high-rate feedback and PODs with $N = M$-dimensional precoders can provide full-diversity. Furthermore, we will show that in an $M$-dimensional MISO system, where $K < M$, which resembles low-rate feedback in our system, the least pairwise error probability is associated to the code structures with $N = K$-dimensional precoders. As a result, to minimize PEP using a generalized POD, we should use $N = \min\{K, M\}$-dimensional precoders.

## C. Pairwise Error Probability Analysis

Using generalized PODs, the equivalent base-band signal at the receiver can be modelled as

$$\mathbf{y} = \mathcal{Z}_j^{\dagger} \; \widehat{\mathbf{h}} \; + \mathbf{n} \tag{5}$$

where $\mathcal{Z}_j^{\dagger}$ is the transmit signal matrix (as the Hermitian of a generalized POD modulation matrix) and $\mathbf{n}$ is the $T$-dimensional noise vector. Note that for simplicity we dropped the upper index of $\mathcal{Z}_j$. If the regulated average SNR at the receiver is $\eta_0$, each element of $\mathbf{n}$ is a complex circularly symmetric additive white Gaussian noise (AWGN) variable with variance $\sigma_n^2 = 1/(M\eta_0)$.

To decode the matrix $\mathcal{Z}_j$, we assume that the index of the constellation set, $j$, which represents the modulation scheme at the transmitter is known at the receiver. Usually, in wireless communications standards, the transmitter sends some control signals to the users (receivers) in the header field of each data frame that indicate the modulation scheme. For example in IEEE 802.16, downlink burst profile of the physical layer, which is a part of downlink channel descriptor of the MAC layer contains the type of modulation used [1]. Our system requires to include the index of the constellation set, $j$, in this category and to assume that the receiver knowledge of $j$ is updated as frequently as feedback is applied. By this assumption, the conditional PEP of detection in favor of an erroneous codeword $\mathcal{Z}_j'$ when $\mathcal{Z}_j$ is transmitted can be tightly upper bounded by the following Chernoff bound [42]

$$\mathrm{p}(\mathcal{Z}_j \to \mathcal{Z}_j' \mid \widehat{\mathbf{h}}) \leq \frac{1}{2} \exp\left( -\frac{D(\mathcal{Z}_j, \mathcal{Z}_j')}{4\sigma_n^2} \right) \tag{6}$$

where $D(\mathcal{Z}_j, \mathcal{Z}_j') = \widehat{\mathbf{h}}^{\dagger} \; (\mathcal{Z}_j - \mathcal{Z}_j')(\mathcal{Z}_j - \mathcal{Z}_j')^{\dagger} \; \widehat{\mathbf{h}}$. Now, suppose that the receiver processes $\overline{h}$ and chooses $\mathcal{V}_i$ from the set of Voronoi regions $\mathcal{V}$. The average probability of pairwise error given the feedback index $i$ at the receiver can be expressed as

$$\text{p}(\mathcal{Z}_j \rightarrow \mathcal{Z}'_j \mid i) = \int_{\text{Dom}(\widehat{\mathbf{h}}|i)} \text{p}(\mathcal{Z}_j \rightarrow \mathcal{Z}'_j|\widehat{\mathbf{h}}) \text{ f}(\widehat{\mathbf{h}}| \ i) \ d\widehat{\mathbf{h}} \tag{7}$$

where $\text{Dom}(\widehat{\mathbf{h}}|i)$ is the domain of the random variable $\widehat{\mathbf{h}}$, conditioned on the receiver's feedback index. The variable $\widehat{\mathbf{h}}$ can be expressed as a one-to-one function of the three variables, $\widetilde{\mathbf{h}}$, $\overline{h}$, and $\gamma$. Therefore, $\text{f}(\widehat{\mathbf{h}}| \ i)d\widehat{\mathbf{h}}$ is statistically equivalent to $\text{f}(\widetilde{\mathbf{h}}; \gamma; \overline{h}| \ i) \ d\widetilde{\mathbf{h}} \ d\gamma \ d\overline{h}$ [34]. Also, for independent and identically distributed (i.i.d.) channel realizations, the latter three variables are independent. Therefore, $\text{f}(\widetilde{\mathbf{h}}; \overline{h}; \gamma| \ i) = \text{f}(\widetilde{\mathbf{h}}) \ \text{f}(\gamma) \ \text{f}(\overline{h}| \ i)$ and we can reexpress (7) as

$$\text{p}(\mathcal{Z}_j \rightarrow \mathcal{Z}'_j \mid i) = \int_{\text{Dom}(\widetilde{\mathbf{h}})} \int_{\mathcal{R}^+} \int_{\mathcal{V}_i} \text{p}(\mathcal{Z}_j \rightarrow \mathcal{Z}'_j| \ \widetilde{\mathbf{h}}, \overline{h}, \gamma) \ \text{f}(\widetilde{\mathbf{h}}) \ \text{f}(\overline{h} \mid i) \ \text{f}(\gamma) \ d\widetilde{\mathbf{h}} \ d\gamma \ d\overline{h} \tag{8}$$

Here, $\mathcal{R}^+ = [0, \infty)$ is the domain of $\gamma$ and $\text{Dom}(\widetilde{\mathbf{h}})$ is the domain of the un-quantized portion of the channel vector. For generalized POD structures, using Equation (6), the above PEP expression can be upper bounded by:

$$\text{p}(\mathcal{Z}_j \rightarrow \mathcal{Z}'_j \mid i) \ \leq \ \frac{1}{2} \int_{\text{Dom}(\widetilde{\mathbf{h}})} d\widetilde{\mathbf{h}} \int_{\mathcal{R}^+} d\gamma \int_{\mathcal{V}_i} d\overline{h} \ \text{f}(\widetilde{\mathbf{h}}) \ \text{f}(\gamma) \ \text{f}(\overline{h}| \ i) \ \exp\left( -\eta_c \left[ \ \gamma \ \beta \ + \ \vartheta \ \right] \right) \tag{9}$$

where $\beta = \overline{h}^\dagger \mathcal{P}_j \mathcal{P}_j^\dagger \overline{h}$ and $\eta_c = \left( \sum_{\kappa=1}^T |z_\kappa - z'_\kappa|^2 \right)/4\sigma_n^2$. Note that $\eta_c$ is proportional to the Euclidian distance of the inner code matrices and the SNR. Also $\vartheta = \left\| \widetilde{\mathbf{h}} \right\|_F^2$ takes values on $[0, \infty)$ and follows a Chi-square distribution. In the sequel, with a slight abuse of the notation, we denote $\mathcal{P}_j^{N \times N}$ by $\mathcal{P}_j$. Because, the precoder dimension $N$ is fixed and the transmitter must pick the precoder matrix from $\mathcal{P} = \{\mathcal{P}_1, \cdots, \mathcal{P}_K\}$ using feedback index $j$. To proceed, we use the following relations:

$$\int_0^\infty \text{f}(\vartheta) \exp\left(-\eta_c\vartheta\right) d\vartheta = (1 + \eta_c)^{-(M-N)}$$

$$\int_{\mathcal{R}^+} \exp\left(-\eta_c \ \gamma \ \beta \ \right) \ \text{f}(\gamma) d\gamma = (1 + \eta_c\beta)^{-N}$$

Then the PEP of the worst-case error event, conditioned on the receiver index $i$ is

$$\text{p}(\mathcal{Z}_j \rightarrow \mathcal{Z}'_j \mid i) \ \leq \frac{1}{2} \ (1 + \eta_c)^{-(M-N)} \ \mathbb{E}_{\overline{h} \in \mathcal{V}_i} \left\{ \ \left( 1 + \eta_c \overline{h}^\dagger \mathcal{P}_j \mathcal{P}_j^\dagger \overline{h} \right)^{-N} \right\} \tag{10}$$

where $\eta_c$ is the minimum Euclidian distance between the inner STBC parts of $\mathcal{Z}_j$ and $\mathcal{Z}'_j$ divided by the noise power. Note that $\eta_c$ represents the transmission signal power divided by the noise power at the receiver antenna or the transmit SNR. When we use unit-norm symbols such as BPSK, in a worst-case error event we have $\eta_c = M\eta_0/(4T)$.

## D. Expanding PEP and Deriving the Design Criterion

Our goal is to minimize the average PEP over all feedback realizations:

$$
\mathrm{P_e} = \sum_{i=1}^{K}\sum_{j=1}^{K} \mathrm{p}(\mathcal{Z}_j \to \mathcal{Z}'_j\,; j\,;\,i) \tag{11}
$$

$$
= \sum_{i=1}^{K}\sum_{j=1}^{K} \mathrm{p}(j\,;\,i)\,\mathrm{p}(\mathcal{Z}_j \to \mathcal{Z}'_j\mid j\,;\,i) \tag{12}
$$

$$
= \sum_{i=1}^{K}\sum_{j=1}^{K} \mathrm{p_f}(j|i)\,\mathrm{p}(i)\,\mathrm{p}(\mathcal{Z}_j \to \mathcal{Z}'_j\mid i) \tag{13}
$$

In the above formulas, the first equality shows the summation of the probabilities that index $i$ is transmitted over the feedback channel, index $j$ is received at the transmitter, and a pairwise error occurs. The second equality is the application of the Bayes' rule. Finally, the third one can be written noting that any two modulation matrices $\mathcal{Z}_j$ and $\mathcal{Z}'_j$ are in the constellation set $j$, chosen given index $j$ at the transmitter. Hence, the event $\mathcal{Z}_j \to \mathcal{Z}'_j\mid i$ is conditioned on $j$.

According to the average PEP expression in (13), the minimum PEP precoder set $\{\mathcal{P}_1, \cdots, \mathcal{P}_K\}$ can be obtained by solving the following optimization problem:

$$
\begin{aligned}
\min \quad & \sum_{j=1}^{K}\sum_{i=1}^{K} \mathrm{p_f}(j|i)\mathrm{p}(i)\, \mathbb{E}_{\overline{h}\in\mathcal{V}_i}\left\{\left(1 + \eta_c\,\overline{h}^\dagger \mathcal{P}_j\mathcal{P}_j^\dagger \overline{h}\right)^{-N}\right\} \\
\text{s.t.} \quad & \forall\, j,\ \ \mathcal{P}_j\mathcal{P}_j^\dagger \succ \mathbf{0}\ \ \text{and}\ \ \mathrm{Tr}(\mathcal{P}_j\mathcal{P}_j^\dagger) = N
\end{aligned} \tag{14}
$$

Here, $\mathcal{P}_j\mathcal{P}_j^\dagger \succ \mathbf{0}$ shows that the Hermitian matrix $\mathcal{P}_j\mathcal{P}_j^\dagger$ is positive semi-definite. We have also omitted the constant term $(1 + \eta_c)^{-(M-N)}$ from the objective function. Equation (14) implies that deriving the set of precoder matrices depends on the feedback indices at the input/output of the quantizer, the set of Voronoi regions $\mathcal{V}$, and the conditional index inversion probability of the feedback channel, $\mathrm{p_f}(j|i)$.

### III. QUANTIZER AND PRECODER CODEBOOK DESIGN

COVQs have been proposed to minimize the quantization average distortion when the quantization indices go through erroneous channels [35]- [38]. We use this idea to design a CSI quantizer for our system. Our COVQ design problem is to find the pair of Voronoi regions and precoder matrices $(\mathcal{V}, \mathcal{P})$ that solves the optimization problem expressed in (14).

First, note that given a fixed set of Voronoi regions $\mathcal{V}$, the joint optimization in (14) can be decoupled into a series of individual optimizations for each $\mathcal{P}_j$ as

$$
\begin{aligned}
\min \quad & \sum_{i=1}^{K} \mathrm{p_f}(j|i)\mathrm{p}(i) \, \mathbb{E}_{\overline{h} \in \mathcal{V}_i} \left\{ \left( 1 + \eta_c \, \overline{h}^\dagger \mathcal{P}_j \mathcal{P}_j^\dagger \overline{h} \right)^{-N} \right\} \\
\text{s.t.} \quad & \mathcal{P}_j \mathcal{P}_j^\dagger \succ \mathbf{0} \ \text{ and } \ \mathrm{Tr}(\mathcal{P}_j \mathcal{P}_j^\dagger) = N
\end{aligned}
\tag{15}
$$

This property simplifies the design procedure significantly. In (15), both the objective function and the constraints are convex functions of $\mathcal{P}_j$. Furthermore, the objective function is differentiable throughout the whole domain of $\mathcal{P}_j$. Therefore, we can use a steepest descent type of algorithm, such as the gradient algorithm to solve this problem [46]. We successively find the Voronoi regions associated to each index $i$ from (14) and the precoder matrices from separate implementations of (15) for every $j$. This algorithm proceeds as follows:

#### A. Training-Based Gradient Algorithm:

1) Generate a sequence of training samples of $\overline{h}$ by normalizing a sequence of $N$-dimensional complex Gaussian random vectors. Then assume an initial set of positive semi-definite precoder matrices $\mathcal{P} = \{\mathcal{P}_1, \cdots, \mathcal{P}_K\}$ with powers $\|\mathcal{P}_j\|_F^2 = N, \ \forall \ j$.

2) Assign index $i$ to each training vector $\overline{h}$ if

$$
i = \arg \min_{\iota \in \{1, \cdots, K\}} \sum_{j=1}^{K} \mathrm{p_f}(j|\iota) \, \left( 1 + \eta_c \, \overline{h}^\dagger \mathcal{P}_j \mathcal{P}_j^\dagger \overline{h} \right)^{-N}
\tag{16}
$$

The set of training vectors with assigned index $i$ statistically represent $\mathcal{V}_i$. Note that the above formula will be used later in the encoder of the quantizer after the codebook $\mathcal{P}$ is designed.

3) To optimize the precoder matrix from (15), first define the following objective function for

each index $j$:

$$\mathbf{J}(\mathcal{P}_j) = \sum_{i=1}^{K} \mathrm{p_f}(j|i) \mathrm{p}(i) \, \mathbb{E}_{\overline{h} \in \mathcal{V}_i} \left\{ \left( 1 + \eta_c \, \overline{h}^\dagger \mathcal{P}_j \mathcal{P}_j^\dagger \overline{h} \right)^{-N} \right\} \tag{17}$$

The gradient of $\mathbf{J}(\mathcal{P_j})$ can be expressed as

$$\nabla \, \mathbf{J}(\mathcal{P}_j) = -2N\eta_c \sum_{i=1}^{K} \mathrm{p_f}(j|i) \mathrm{p}(i) \, \mathbb{E}_{\overline{h} \in \mathcal{V}_i} \left\{ \left( 1 + \eta_c \, \overline{h}^\dagger \mathcal{P}_j \mathcal{P}_j^\dagger \overline{h} \right)^{-N-1} \, \overline{h} \, \overline{h}^\dagger \, \mathcal{P}_j \right\} \tag{18}$$

In numerical implementation of (18), note that the random vector $\overline{h}$ is uniformly distributed on the region $\mathcal{V}_i$ and its pdf is proportional to the volume of $\mathcal{V}_i$, or the marginal probability $\mathrm{p}(i)$. For any function $\Psi$, to find $\mathrm{p}(i) \, \mathbb{E}_{\overline{h} \in \mathcal{V}_i} \{\Psi(\overline{h})\}$, it is sufficient to add the values of $\Psi(\overline{h})$ throughout the partition of the training space that is represented by index $i$ (approximation of $\mathcal{V}_i$) and divide the result by the size of the training sequence.

4) To proceed with the gradient algorithm, update each matrix in the precoder codebook $\mathcal{P}$ using the following relation:

$$\mathcal{P}_j(t+1) = \left[ \begin{array}{c} \mathcal{P}_j(t) \, - \, \alpha(t) \, \nabla \mathbf{J}(\mathcal{P}_j(t)) \end{array} \right]_N^+ \tag{19}$$

In the above formula, $\mathcal{P}_j(t)$ denotes the value of the precoder matrix $\mathcal{P}_j$ in the $t$-th iteration of the algorithm. The step size $\alpha(t)$ can be set to any decreasing function of $t$, but from [46] we use $(1+m)/(1+t)$ with an arbitrary positive real number $m$. The operation $[\,\cdot\,]_N^+$ shows that the matrix in the brackets is projected onto the space of positive semi-definite matrices with power $N$. We use the Euclidean distance as the projection measure. With this measure, the above projection is simply removing all the negative eigenvalues of the matrix inside the brackets and normalizing the matrix, so that its power is $N$ [47].

5) Use the output of the above algorithm and improve it by successively implementing steps (2) and (3-4) until convergence. Since the resulting sequence of objective function values from (14) is decreasing and PEP is bounded bellow, the convergence of this algorithm is guaranteed. However, we cannot claim that by the above alternation we converge to a globally optimal solution.

After storing the precoder matrix codebook $\mathcal{P}$ at the transmitter/receiver ends of the system, the encoder of the quantizer at the receiver side operates similar to Equation (16). At the transmitter side, the task of the quantizer decoder is to choose $\mathcal{P}_j$ from the precoder codebook.

In general, we can implement different index mapping methods in our scheme. For incorporating a mapping other than identity, we map the output index of the quantizer $i$ to $i'$ at the receiver side and pass it through the feedback channel. Then at the transmitter side, we use inverse mapping (demapping) to obtain the quantization index $j$ from the output of the feedback channel, i.e., $j'$. Note that to drive the system with different mappings, one should replace $\mathrm{p_f}(j|i)$ with $\mathrm{p_f}(j'|i')$ in the design of the precoder codebook and in the implementation of the quantizer's encoder.

### B. System Design Without Knowing $\rho_f$ A Priori

The design techniques established in the previous section can be extended to the case that the knowledge of $\rho_f$ is not accurate. Suppose that we know a coarse range of feedback bit error probability $\rho_f$ at the transmitter/receiver sides. The uncertainty about the feedback channel model can be taken into account assuming that $f_a \leq \rho_f \leq f_b$, where $f_a, f_b \in [0, 0.5]$. This type of channel modeling mismatch results from the uncertainty about the feedback channel conditions in a wireless environment and the uncertainty about the amount of resources that different receivers in the network may spend for protecting feedback signals.

We develop our design strategy based on assuming a fixed $\rho_d \in [f_a, f_b]$, which is called the design cross-over probability parameter. Inspired from [39], it is known that by designing COVQs based on worst-case assumptions, the system enjoys desirable robustness against channel modeling mismatch. Therefore, we find the set of precoder matrices and Voronoi regions that minimize the worst possible pairwise error probability. Similar to (14), the precoder design criterion can be expressed as:

$$\min_{\rho_d \in [f_a, f_b]} \max \quad \sum_{j=1}^{K} \sum_{i=1}^{K} \mathrm{p_d}(j|i)\mathrm{p}(i) \, \mathbb{E}_{\overline{h} \in \mathcal{V}_i} \left\{ \left( 1 + \eta_c \, \overline{h}^\dagger \mathcal{P}_j \mathcal{P}_j^\dagger \overline{h} \right)^{-N} \right\} \tag{20}$$
$$\text{s.t. } \forall j, \ \mathcal{P}_j \mathcal{P}_j^\dagger \succ \mathbf{0} \ \text{ and } \ \mathrm{Tr}(\mathcal{P}_j \mathcal{P}_j^\dagger) = N$$

where $\mathrm{p_d}(j|i)$ is the conditional index inversion probability between $i$ and $j$ assuming that the BSC cross-over probability is $\rho_d$. It is straightforward to show that the objective function in (20) is an increasing function of the design cross-over probability parameter $\rho_d$. We can show this property by plotting $P_e$ from (13) versus the SNR. Therefore, the $\min\max$ value of $\rho_d$ coincides with the maximum point of the cross-over probability range, which is $\rho_d = \max\rho_f = f_b$. By this choice, the rest of the COVQ design procedure from Section III-A can be applied.

## IV. NUMERICAL RESULTS

### A. Precoder Characteristics

In this section, we study the precoder structure of a 4-antenna system ($M = 4$) with 16 feedback indices ($K = 16$) or 4 feedback bits ($\log_2 K = 4$). The precoder codebook is designed for different cross-over probabilities $\rho_f$. The solutions are obtained by solving optimization problem (14), using the training based gradient algorithm explained in Section III-A. Fig. 2 depicts the eigenvalues of the first member of the precoder codebook, $\mathcal{P}_1$, i.e., $[\delta_1, \cdots, \delta_4]$. Other members of the codebook also have similar properties. Here, $\mathcal{P}_1\mathcal{P}_1^\dagger = \mathcal{U}_1 \operatorname{Diag}[\delta_1^2, \cdots, \delta_4^2]\, \mathcal{U}_1^\dagger$, where $\mathcal{U}_1$ is unitary and $\operatorname{Diag}[\cdot]$ denotes a diagonal matrix. Note that the transmitter's power constraint requires that $\sum_{\kappa=1}^4 \delta_\kappa^2 = 4$. When the feedback link is error-free or $\rho_f = 0$, all the transmission power is assigned to the first eigenmode, denoted by the largest eigenvalue, $\delta_1$. In other words, to achieve minimum PEP, transmission power is projected onto the direction of the major eigenvector of the precoder matrix, i.e. a column of $\mathcal{U}_1$. This result coincides with the optimality of directional beamforming. With an error-free feedback link, we can design a directional beamforming system, where the transmit signals are the product of scalar symbols and unit-norm beamforming vectors. Hence, our precoding system degenerates to a beamforming system, where the first eigenvector of the precoder matrix is equivalent to the beamforming direction. As the cross-over probability of the feedback link increases, the eigenvalues of the minimal PEP precoder matrix converge to equal values. A better error performance can be obtained by spreading the power among different directions in space. In this case, our precoded matrix constellations also converge to matrices with equal eigenvalue squares, similar to open-loop orthogonal space-time block codes.

### B. BER Computations

Fig. 3 shows the Bit Error Rate (BER) of our combining scheme over a $4{\times}1$ MISO channel using Monte Carlo simulations. The feedback link carries 4 bits with different cross-over probabilities.

This situation resembles a high-rate feedback assumption $K \geq M$. The transmitter uses BPSK constellation symbols and the transmission rate is 1 bit/sec/Hz. In each data frame we transmit 130 data symbols. In this figure, we demonstrate the performance of an open-loop STBC and that of the combining scheme without feedback errors. By increasing the cross-over probability $\rho_f$ from 0.00 to 0.50, the BER of the system ranges between a directional beamforming system and an open-loop system. All the curves show full-diversity order within the range of SNRs that we consider in our simulations. This figure shows that even with feedback errors, our combining scheme preserves full-diversity order and it also provides additional array gain compared to an open-loop STBC. In this figure, we also plot the BER of Grassmannian beamforming from [15] [25] and show that it is optimal in terms of minimizing the PEP when feedback is error-free. According to our numerical experiments, different index mapping/demapping schemes at the input/output of the feedback channel result in similar performance results if the cross-over probability of the feedback link is known and the precoder codebook is optimized for each specific mapping.

The BER performance of our scheme with low-rate and also erroneous feedback is illustrated in Fig. 4. In this experiment, we consider a $6 \times 1$ MISO channel with 2 feedback bits or 4 feedback regions. Directional beamforming in this case cannot achieve full-diversity order. Our combining system also shows the same property if we use 6-dimensional precoder matrices. In order to obtain full-diversity order, PSTBCs in the literature for instance the unitary PSTBCs in [25] use $6 \times 2$ precoder matrices from unitary constellations in [48], applied to $2 \times T$ orthogonal design matrices [42]. We use generalized POD constellation matrices similar to the one introduced in (3), based on a $6 \times 8$ orthogonal code structure. This code is generated by removing two rows of an $8 \times 8$ orthogonal design, leaving the first two rows without precoding, and multiplying $4 \times 4$ precoder matrices inside the code structures at the lower 4 rows of the matrix. The simulation curves can be extended using other values of $N$, which we skip for the sake of clarity in the figure. Our conclusion from this study is that only when $N \leq K$, full diversity order can be obtained. Moreover, when $K < M$, minimum BER performance is attributed to the POD structure with $K$-dimensional precoders. Over an error-free feedback channel, our combining scheme with 4-dimensional precoders outperforms PSTBCs with $6 \times 2$ unitary precoders. Fig. 4 also shows the performance of the system with different feedback cross-over probabilities. With a non-zero cross-over probability in the feedback link, PSTBCs with unitary precoders fail to obtain full-diversity order. Our combining scheme with 4-dimensional precoder matrices, designed based on knowing $\rho_f$, however preserves full-diversity order, even if feedback is low-rate and erroneous.

In Fig. 5, we show the performance of our combining schemes over a 4-antenna MISO channel with complex constellations. There is no $M \times T$ rate-one complex orthogonal design for $M > 2$ [42]. In this case, instead of using orthogonal design structures with less rates, we can use a quasi-orthogonal space-time block code (QOSTBC) [43] [44], to obtain a generalized partly quasi-orthogonal design (PQOD) structure. The quasi-orthogonal inner code of this example uses QPSK modulation symbols with $\frac{\pi}{4}$ rotations. The transmission rate is 2 bits/sec/Hz. With $K = 4$ and $K = 2$ feedback regions, we use generalized PQODs with $N = 4$ and $N = 2$-dimensional precoders, respectively. Again, with feedback errors, these constellations can provide full-diversity order and additional array gain compared to open-loop QOSTBCs. In this figure, we also demonstrate the performance of this system employing Grassmannian beamforming, when feedback is erroneous. This situation resembles a noiseless design mismatch and results in not achieving full-diversity order. Here, the number of feedback regions is equal to the number of transmit antennas and Grassmannian beamforming is similar to antenna selection (AS). The performance of our combining scheme designed assuming noiseless feedback (VQ-design) is also very close to that of Grassmannian beamforming. In this case, there are $N$ independent beamforming directions in space with similar cross (Chordal) distances [15], [25]. Upon receiving an erroneous feedback index, it does not matter which direction we pick. Hence, employing different index mapping schemes in this case does not change the performance of the system, even when the feedback channel is erroneous.

## C. Effect of Mismatch

In the last experiment, we investigate the effects of channel modeling mismatch on the performance of our system. For the numerical results shown in Fig. 6, we again consider a $4 \times 1$ MISO channel with 16 feedback regions and BPSK symbol constellations. We assume certain feedback cross-over probabilities for designing the COVQ or VQ-based combining system and use different ones for Monte Carlo simulations. The following cases are studied:

First, the precoder codebook is designed assuming feedback is noiseless. We call this design a VQ-based one since COVQ becomes a VQ when $\rho_d = 0.00$. Also the combining scheme degenerates to directional beamforming. The resulting system is examined using a feedback link, where the actual cross-over probability is $\rho_f = 0.04$. The performance results are very close to the performance of Grassmannian beamforming when operating over an erroneous feedback channel with $\rho_f = 0.04$. This type of mismatch, which results in severe performance degradation is referred to as VQ-based design mismatch. In this case, since the number of feedback regions is more than the number of transmit antennas, different beamforming directions have different cross (Chordal) distances [15], [25]. Hence, we can improve the performance of the system by modifying the feedback index mapping without any additional cost. To show this property, we first define the average Chordal distance of the beamforming directions at the transmitter, conditioned on the feedback indices at the receiver, when the feedback cross-over probability is given. Then we minimize this average distance over the space of possible index mapping solutions using simulated annealing (SA) [37]. The details of this algorithm are skipped for brevity. The resulting VQ-based system outperforms the same system with identity mapping. However, the gain of optimizing the index mapping is not significant compared to the gain of COVQ-based design since it does not change the diversity of the system. Second, we assume the design cross-over probability $\rho_d = 0.04$ and examine the system performance over an error-free feedback link, where $\rho_f = 0.00$. This situation is called COVQ-based design mismatch. This type of mismatch does not degrade the performance of the system severely. The performance results in this case are worse than those of a system with ideal design of precoders ($\rho_d = 0.00$) and error-free simulations ($\rho_f = 0.00$). However, full-diversity order is still preserved.

Third, we consider a feedback channel where the cross-over probability changes uniformly in the range $0.00 \leq \rho_f \leq 0.04$. In this scenario, as we explained in Section III-B, we use a worst-case design strategy to obtain robustness against feedback channel modeling mismatch. The precoder codebook is designed assuming that the cross-over probability is $\rho_d = 0.04$. By this choice, full-diversity order is preserved and also the array-gain of the system is superior to an open-loop system. Fourth, we consider the same feedback channel model with the same range of cross-over probabilities, but this time we assume an alternative design parameter, where $\rho_d$ is the average value of feedback errors, i.e., $\rho_d = (f_a + f_b)/2 = 0.02$. By this assumption, the transmission scheme does not achieve full-diversity order. The latter observation confirms that the worst case design assumption $\rho_d = \max_{[f_a, f_b]} \rho_f$ is a better candidate compared to the average error design.

## V. Summary and Conclusions

In this paper, we studied high-rate and low-rate feedback (closed-loop) communication systems with possible feedback errors. We showed numerically that our combining schemes obtain full-diversity order with additional array gain compared to open-loop systems. The design criterion of our scheme was derived based on a pairwise error probability measure and our system was optimized using a training-based algorithm. As the feedback cross-over probability approaches zero, our scheme converges to a directional beamforming system. On the other hand, as the feedback error increases, our scheme converges to a no-CSIT or open-loop system. We presented combining strategies for both high-rate and very low-rate feedback scenarios, which is simply extendable to the cases that the feedback link is noisy. With very low-rate feedback, the dimension of the precoder matrix in our scheme reduces and our system converges to an open-loop STBC structure.

The design strategies and solution algorithms presented in this paper are robust against feedback channel modeling mismatch. Using a worst-case design approach, we designed the quantizer structure and derived the precoder matrices used in the code structures, even if the exact value of the feedback channel error is unknown to the transmitter/receiver sides a priori. It was shown numerically that even in this case, our transmission scheme preserves full-diversity order and provides additional array gain compared to open-loop systems.

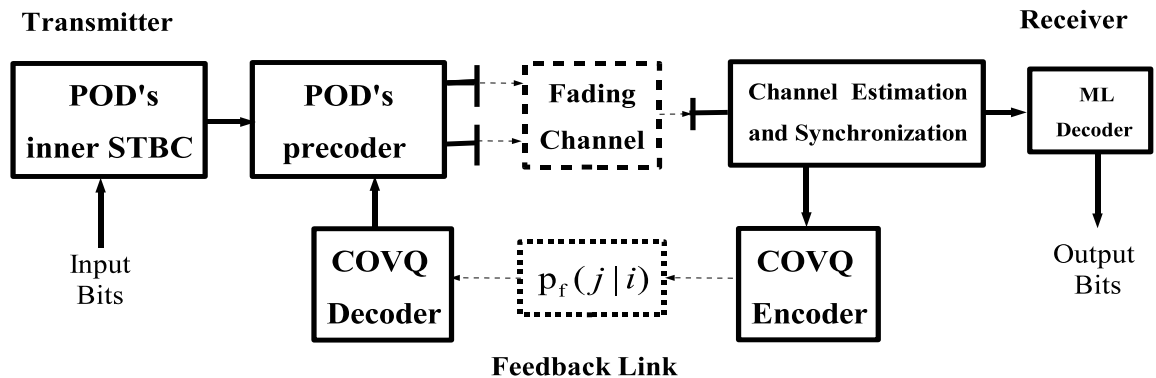

Fig. 1.   System block diagram.

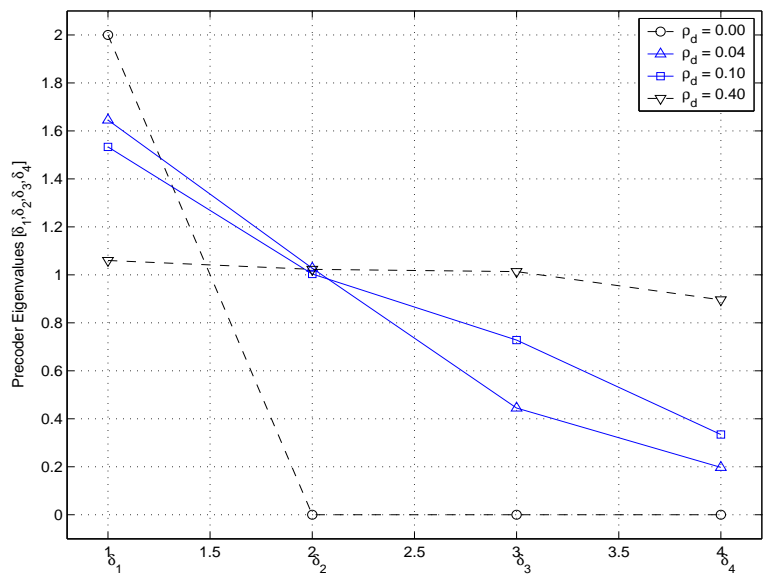

Fig. 2.   Eigenvalues of $\mathcal{P}_1$, $[\delta_1, \cdots, \delta_4]$ for an $M = 4$ multi-antenna system with $K = 16$ feedback regions and different cross-over probabilities.

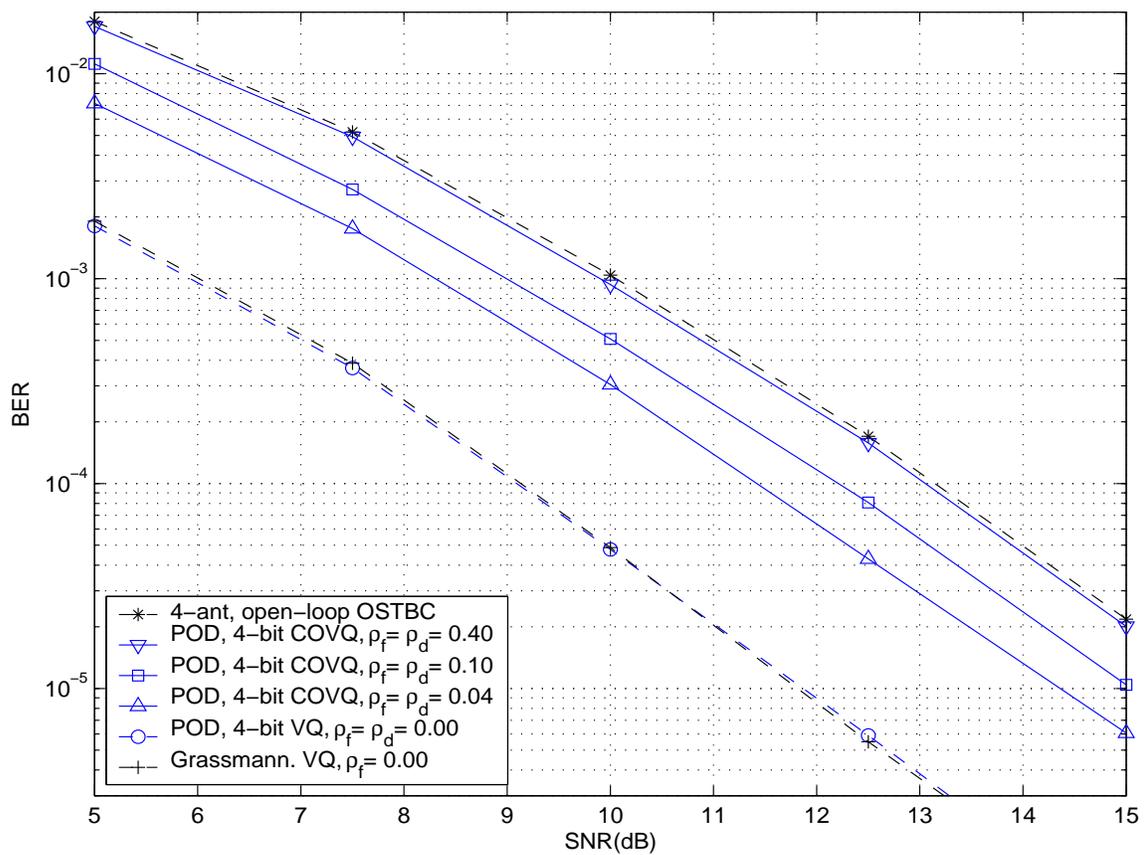

Fig. 3.  Bit Error Rate for 4-ant. generalized PODs with 4 bits per vector COVQs. Rate = 1 bit/sec/Hz using BPSK symbols.

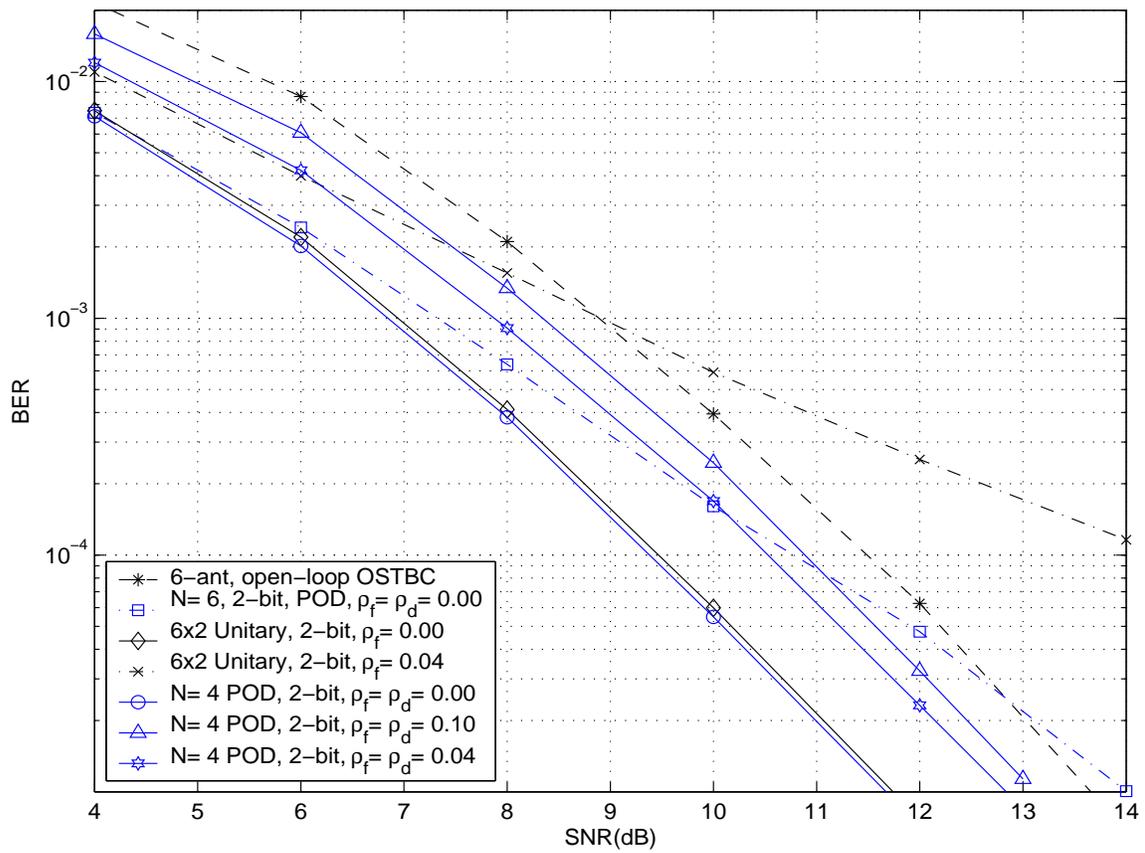

Fig. 4. Bit Error Rate for 6-ant. generalized PODs with 2 bits per vector COVQs. Rate = 1 bit/sec/Hz using BPSK symbols.

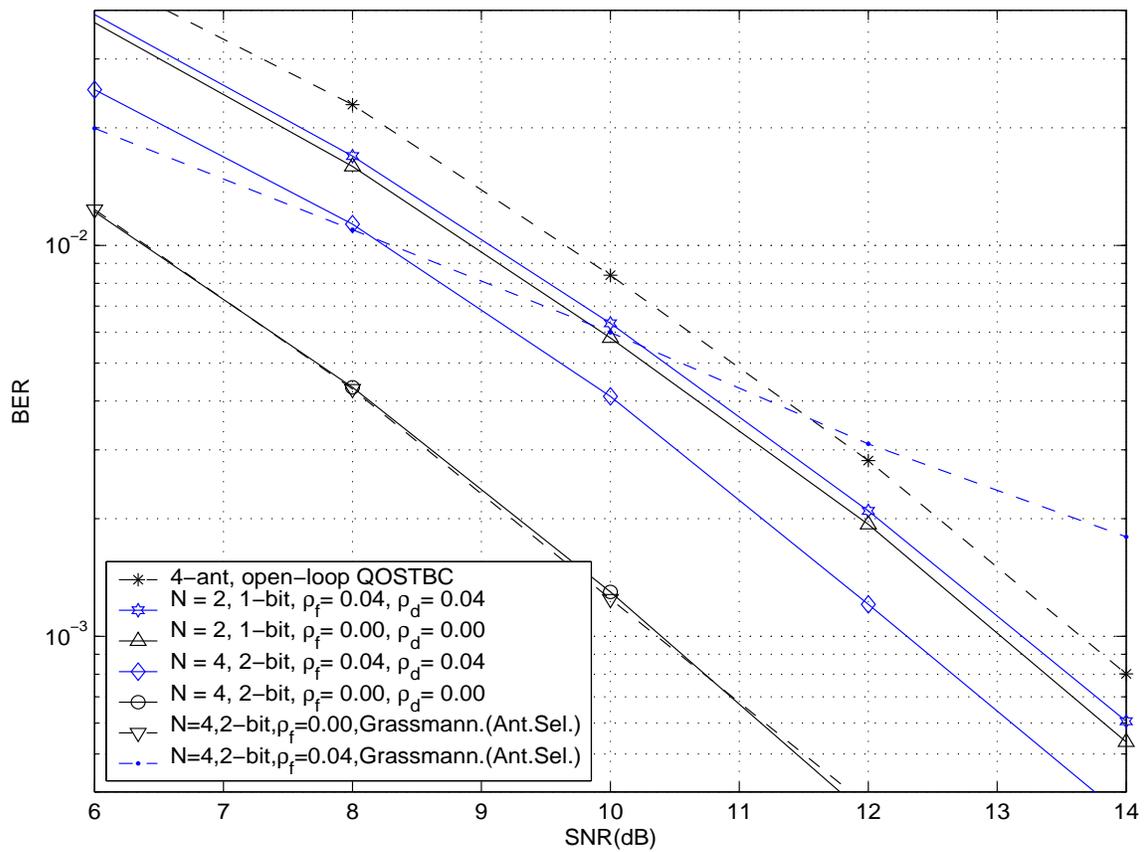

Fig. 5.  Bit Error Rate for 4-ant. generalized PQODs with 2 bits and 1 bit per vector COVQs. Rate = 2 bits/sec/Hz using QPSK symbol constellations with $\pi/4$ rotations.

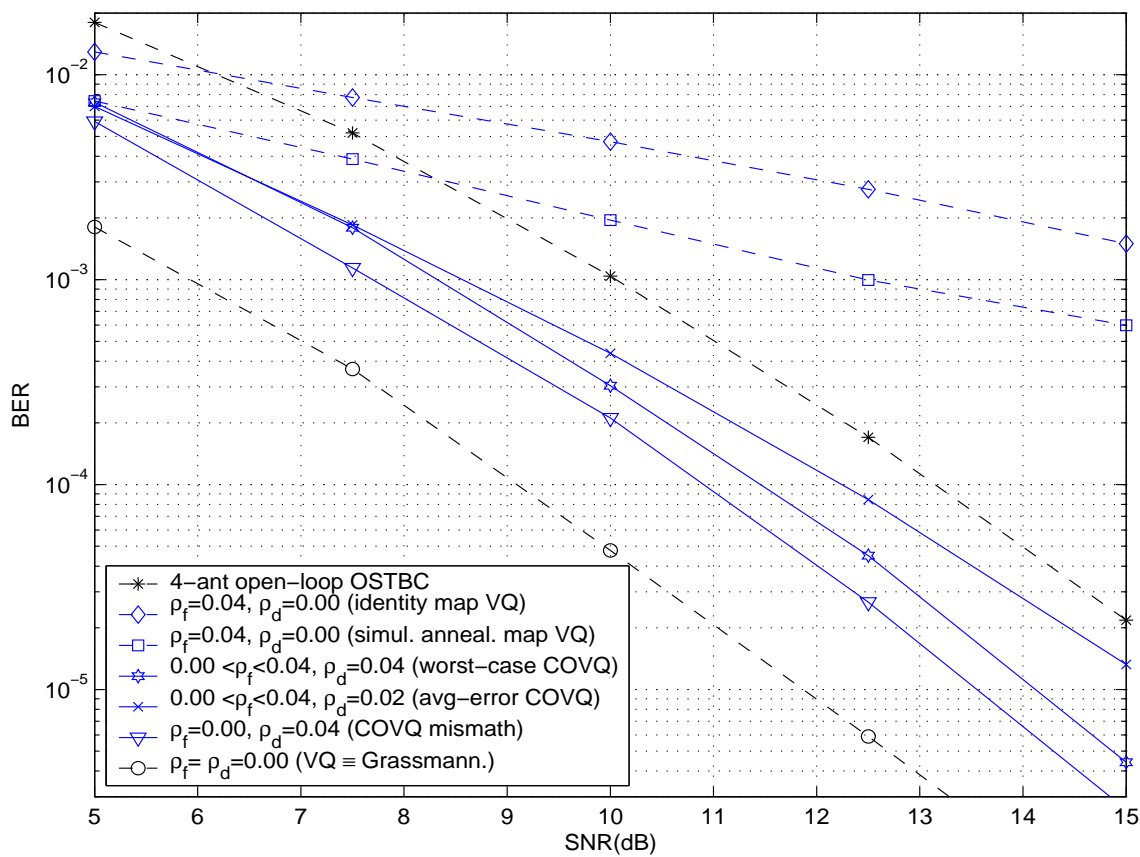

Fig. 6. Bit Error Rate for 4-ant. generalized PODs with 4 bits per vector COVQs. Rate = 1 bit/sec/Hz using BPSK symbols with feedback channel modeling mismatch, ($\rho_d \neq \rho_f$).